\documentclass[a4paper]{article}
\usepackage{spconf}

\usepackage[utf8]{inputenc}

\usepackage{color}

\usepackage{amsmath,amssymb,bm}
\usepackage{cite}

\usepackage[acronyms]{glossaries}
\makeglossaries

\usepackage{pgfplots}
\pgfplotsset{compat=1.5}
\usepackage{tikz}
\usetikzlibrary{decorations,calligraphy,calc}

\usepackage[hidelinks]{hyperref}

\newcommand{\exportFigures}{false}
\newcommand{\exportFiguresAsPNG}{false}

\ifthenelse{\equal{\exportFigures}{true}}
{
	\usepgfplotslibrary{external}
	\tikzexternalize[prefix=compiled_tikz_figures/]
	\ifthenelse{\equal{\exportFiguresAsPNG}{true}}
	{
		\tikzset
		{   png export/.style={
				external/system call={
					pdflatex \tikzexternalcheckshellescape -halt-on-error --extra-mem-top=10000000 -interaction=batchmode -jobname "\image" "\texsource" && pdftops -eps "\image.pdf" && convert -density 700 -transparent white "\image.pdf" "\image.png"
		}}}
		\tikzset{png export}
	}
	{}
}
{}

\newlength{\figurewidth}
\newlength{\figureheight}

\title{\Large Variational Message Passing-based Respiratory Motion Estimation and Detection Using Radar Signals}

\name{Jakob Möderl, Erik Leitinger, Franz Pernkopf, and Klaus Witrisal
	\thanks{This research was partly funded by the Austrian Research
		Promotion Agency (FFG) within the project SEAMAL Front (project
		number: 880598).}}
\address{\small Graz University of Technology, Graz, Austria (jakob.moederl@tugraz.at)}

\newcommand{\Tst}{T_{\text{rep}}}
\newcommand{\fc}{f_{\text{c}}}

\newcommand{\iu}{j}

\newcommand{\trace}{\text{tr}}

\newcommand{\real}{\text{Re}}

\newcommand{\transpose}{{\text{T}}}
\newcommand{\hermitian}{{\text{H}}}

\newcommand{\ComplexNums}{\mathbb{C}}

\newcommand{\normpdf}[3]{\mathcal{N}\left.\big(#1\,\big|\,#2,\,#3 \big)\right.}
\newcommand{\cnormpdf}[3]{\mathcal{C}\normpdf{#1}{#2}{#3}}

\newacronym{snr}{SNR}{signal-to-noise ratio}
\newacronym{fft}{FFT}{fast Fourier transform}
\newacronym{uwb}{UWB}{ultra-wideband}
\newacronym{vmp}{VMP}{variational message passing}
\newacronym{elbo}{ELBO}{evidence lower bound}
\newacronym{los}{LoS}{line-of-sight}
\newacronym{mpc}{MPC}{multipath component}

\providecommand{\rmv}{\hspace*{-.3mm}}

\newcommand{\SecHeadSpacingPre}{-3mm}
\newcommand{\SecHeadSpacingPost}{-2mm}

\begin{document}
\maketitle


\begin{abstract}
We present a \gls{vmp} approach to detect the presence of a person based on their respiratory chest motion using multistatic \gls{uwb} radar.
In the process, the respiratory motion is estimated for contact-free vital sign monitoring.
The received signal is modeled by a backscatter channel and the respiratory motion and propagation channels are estimated using \gls{vmp}.
We use the \gls{elbo} to approximate the model evidence for the detection.
Numerical analyses and measurements demonstrate that the proposed method leads to a significant improvement in the detection performance compared to a \gls{fft}-based detector or an estimator-correlator, since the \glspl{mpc} are better incorporated into the detection procedure.
Specifically, the proposed method has a detection probability of $0.95$ at $-20$\,dB \gls{snr}, while the estimator-correlator and \gls{fft}-based detector have $0.32$ and $0.05$, respectively.
\end{abstract}


\glsresetall

\vspace{\SecHeadSpacingPre}
\section{Introduction}
\vspace{\SecHeadSpacingPost}
%

Future cars will be required to detect if they are occupied when the car is being locked to prevent the confinement of small children, which makes detecting the presence of people a safety-critical application \cite{euroncapRoadmap2025,qinyi2020, aghaei2016}.
The respiratory motion of the chest provides a dynamic feature used to separate the target's radar response from the strong clutter present in this use case \cite{moederl22}.
We derive a Kronecker-factorized signal model for vital sign estimation in (strong) clutter using \gls{uwb} radar and present a \gls{vmp} algorithm \cite{winn2005,bishop2009ApproxInf} to detect the presence of people and estimate the respiratory chest motion. An advantage of the derived model is, that it is linear in both the respiratory motion as well as the propagation channel and, thus, the message passing equations can be solved analytically.


Recent approaches to detect and estimate the respiratory chest motion based on radar responses are mostly based on the intuition that the respiratory motion is periodic. They apply techniques such as \gls{fft} processing, principal component analysis or energy detection \cite{baird17,ahmad18,anitori09,kilic14,yang21}. 
However, this assumption is easily violated, e.g. by infants who regularly experience apnea (short pauses with no respiration) \cite{frey98}. This is especially important, considering that infants are very hard to detect in the first place, due to the small radar cross section and respiratory motion amplitude.
Other works such as \cite{ma20,alizadeh19:IEEEsensors} rely on a data-driven approach, which is typically limited by the small and heterogeneous data sets available.
None of these works explicitly models the propagation of \glspl{mpc}. However, in a tightly enclosed space, such as the interior of a car, the \glspl{mpc} that interact with the target carry additional information and can be used to increase the \gls{snr}, which is of critical importance in the given use case. In this work, we apply \gls{vmp} to improve upon the results of \cite{moederl22}, which already incorporates \glspl{mpc} into the detection.

\emph{Notation}:
We define $\bm{I}_n$ to be the $n\times n$ identity matrix and $\bm{1}=[1,\,1,\,\cdots,\,1]^\transpose$ to be a vector of ones with appropriate length.
We use $\odot$ and $\otimes$ to denote the Hadamard (element wise) and Kronecker product of two vectors or matrices, respectively.
The real operator and matrix trace operator are denoted as $\real\{\cdot\}$ and $\trace(\cdot)$. We use $\mathcal{N}(\bm{a}|\bm{b},\,\bm{C})$ and $\mathcal{CN}(\bm{a}|\bm{b},\,\bm{C})$ to denote that the vector $\bm{a}$ is distributed with a (complex) multivariate Gaussian distribution with mean $\bm{b}$ and covariance matrix $\bm{C}$.
Similarly, $\text{Ga}(a|d,\,e)$ is used to denote that the variable $a$ is gamma-distributed with shape parameter $d$ and rate parameter $e$.
The differential entropy of the distribution $q(x)$ is denoted as $\mathbb{H}(p)$ and the expectation of the function $f(x)$ with respect to $q(x)$ as $\big<f(x)\big>_{q(x)}$.

\vspace{\SecHeadSpacingPre}
\section{Signal Model}
\label{sec:signal_model}
\vspace{\SecHeadSpacingPost}

We consider the case of a person sitting in a car, without intentional body movement. However, the chest of the person expands and contracts continuously due to the persons respiration. We propose to model 
the chest movement in direction of the antenna $b_{\text{t}}(t)$ as the realization of a zero-mean Gaussian random process. We aim to detect the presence of the person and estimate $b_{\text{t}}(t)$ using multistatic \gls{uwb} radar signals.

Let $\bm{s}$ be $N$ samples, equally spaced with spacing $\Delta f=f_{\text{s}}/N$, of the complex baseband representation of the transmit pulse $s(f)$ centred at carrier frequency $\fc$. Several repetitions of the pulse are transmitted at times $t=mT_{\text{rep}}$, $m\in\left\{0,\,1,\,\cdots,\,M-1\right\}$ from the transmit antenna.
After propagating over $K$ time-varying channels with frequency response $\bm{h}_{k}(t)$, each signal $\bm{r}_{k,m} = \left.\bm{h}_{k}(m\Tst)\odot\bm{s}\right. + \bm{w}_{k,m}$
received at receiving antenna $k$ at repetition $m$ is corrupted by noise $\bm{w}_{k,m}$.
The noise samples $\bm{w}_{k,m}$ are generated by a noise process $W_{m,k}(f)$, which is modeled as additive white Gaussian noise with double-sided power spectral density $N_0/2$, and is assumed to be independent across $m$, $k$ and $f$. Thus, $\bm{w}_{k,m}$ is a circular symmetric complex Gaussian random vector with covariance $\bm{C}_{\text{w}}=\lambda^{-1}\bm{I}_{N}$ and precision $\lambda=T_{\text{s}}/N_0$.

In order to remove the clutter, the signals from all antennas at time $m$ are stacked $\bm{r}_{m}=[\bm{r}_{1,m}^\transpose,\,\bm{r}_{2,m}^\transpose,\,\cdots,\,\bm{r}_{K,m}^\transpose]^\transpose$ and the mean $\bar{\bm{r}}=\frac{1}{M} \sum_{m=0}^{M-1} \bm{r}_{m}$ over $m$ is subtracted $\tilde{\bm{r}}_{m}=\bm{r}_{m}-\bar{\bm{r}}$. Finally, the signals are stacked into a large column vector $\tilde{\bm{r}}=\left.[\tilde{\bm{r}}_{0}^\transpose,\,\tilde{\bm{r}}_{1}^\transpose,\,\cdots,\,\tilde{\bm{r}}_{M-1}^\transpose]^\transpose\right.$. 
As we derive in the following subsection, the time-varying part of the received signal 
\vspace{-3mm}%
\begin{equation}
	\tilde{\bm{r}} = \bm{b}_{\text{t}}\otimes\bm{h}_{\text{s}} + \bm{w}
	\label{eq:signal_model_kron}
	\vspace{-1mm}%
\end{equation}
is given as the product of the respiratory motion $\bm{b}_{\text{t}} = \big[b_{\text{t}}(0),\,b_{\text{t}}(T_{\text{rep}}),\,\cdots,\,b_{\text{t}}\big((M-1)T_{\text{rep}}\big)\big]^\transpose \in \mathbb{R}^{M}$ and a stacked channel vector $\bm{h}_{\text{s}} = [\bm{h}_{\text{s},1}^\transpose,\,\bm{h}_{\text{s},2}^\transpose,\cdots,\,\bm{h}_{\text{s},K}^\transpose]^\transpose$ in additive white Gaussian noise $\bm{w} = [\bm{w}_{1,0}^\transpose,\, \bm{w}_{2,0}^\transpose,\, \cdots,\, \bm{w}_{K,M-1}^\transpose]^\transpose$.

Since we apply a frequency selective prior to $\bm{b}_{\text{t}}$ the resulting covariance $\bm{C}_{\text{b}_{\text{t}}}$ is not full rank. Thus, all computations are performed in the eigenspace $\bm{b} = \bm{U}^{\text{T}}\bm{b}_{\text{t}}$ corresponding to the eigendecomposition $\bm{C}_{\text{b}_{\text{t}}}=\bm{U}\bm{C}_{\text{b}}\bm{U}^{\text{T}}$, where $\bm{C}_{\text{b}}$ is a diagonal matrix with all $L$ non-zero eigenvalues of $\bm{C}_{\text{b}_{\text{t}}}$ on its main diagonal and $\bm{U}$ is a matrix with the corresponding eigenvectors.
Let $\bm{e}_k = [0,\,\cdots,\,0,\,1,\,0,\,\cdots,\,0]^\transpose$ be a vector of length $K$ with all zeros except for an 1 at the $k$-th position and $\bm{w}_k = [\bm{w}_{k,0}^\transpose,\,\bm{w}_{k,1}^\transpose,\,\cdots,\,\bm{w}_{k,M-1}^\transpose]^\transpose$, $\tilde{\bm{r}}$ can be expressed either as a linear function of the breathing signal $\bm{b}_{\text{t}}=\bm{U}\bm{b}$ given the block-diagonal matrix $\bm{H}=\bm{I}_{M}\otimes\bm{h}_{\text{s}}$ and the signal received at each antenna $k$ after clutter removal $\tilde{\bm{r}}_{\text{A}k}=\big((\bm{e}_k^\transpose\otimes\bm{I}_N\big)\otimes\bm{I}_M)\tilde{\bm{r}}$ as a linear function of $\bm{h}_{\text{s},k}$ given the block-diagonal matrix $\bm{B}=\bm{U}\bm{b}\otimes\bm{I}_{N}$:
\vspace{-2mm}%
\begin{gather}
	\tilde{\bm{r}} =\bm{H}\bm{U}\bm{b}+\bm{w} 	\label{eq:signal_model} \\	
	\tilde{\bm{r}}_{\text{A}k} = \bm{B}\bm{h}_{\text{s},k}+\bm{w}_{k} \nonumber
	.
	\vspace{-2mm}%
\end{gather}
Assuming independent channels $\bm{h}_{\text{s},k}$, the likelihood of receiving $\tilde{\bm{r}}$ is $p(\tilde{\bm{r}}|\bm{b},\bm{h}_{\text{s}},\lambda)=\cnormpdf{\tilde{\bm{r}}}{\bm{H}\bm{U}\bm{b}}{\lambda^{-1}\bm{I}_{KNM}} = \prod_{k=1}^{K} \cnormpdf{\tilde{\bm{r}}_{\text{A}k}}{\bm{B}\bm{h}_{\text{s},k}}{\lambda^{-1}\bm{I}_{NM}}$.

\vspace{\SecHeadSpacingPre}
\subsection{Propagation environment and target model}
The propagation environment is modeled as a time-varying backscatter channel \cite{arnitz12} with frequency response
\vspace{-2mm}%
\begin{equation}
	\bm{h}_k(t)=\bm{h}_{\text{t},k}(t) \odot \bm{h}_{\text{fb},k} + \bm{h}_{\text{c},k}
	\label{eq:backscatter-channel}
	\vspace{-2mm}
\end{equation}
for each receiving antenna.
Introducing briefly the target channel $\bm{h}_{\text{t},k}(t)$,
the \glspl{mpc} which interact with the target are modeled by a forward-backward channel $\bm{h}_{\text{fb},k}=\bm{h}_{\text{f},k}\odot\bm{h}_{\text{b},k}$, which is the product of a forward channel $\bm{h}_{\text{f},k}$ covering the propagation from the transmit antenna to the target and a backward channel $\bm{h}_{\text{b},k}$ covering the propagation from the target back to the receive antenna. All other received \glspl{mpc} are termed as clutter and are modeled by the frequency response $\bm{h}_{\text{c},k}$. Since the target is assumed to be stationary, the channels $\bm{h}_{\text{f},k}$, $\bm{h}_{\text{b},k}$ and $\bm{h}_{\text{c},k}$ can be assumed time-invariant as long as the respiratory motion $b_{\text{t}}(t)$ is much smaller than the smallest wavelength of the transmit signal.

The target is modeled as a single point target with a time-varying baseband frequency response $h_{\text{t},k}(f,t)=\alpha_k e^{-\iu2\pi (f+\fc)\tau_{\text{b},k}(t)}$, representing the reflection of the incoming signal by a coefficient $\alpha_k \in \ComplexNums$ and a time-varying delay $\tau_{\text{b},k}(t)$. Let $c$ be the propagation speed of the signal and $0 < \rho_k < 2$ a coefficient depending on the angles between the target and the transmitting and receiving antennas, $\tau_{\text{b},k}(t) = \rho_k\, b(t) / c$.
Using a first order Taylor-approximation $e^{-\iu2\pi (f+\fc)\tau_{\text{b}}(t)} \approx 1-\iu 2\pi (f+\fc)\tau_{\text{b},k}(t)$, the sampled frequency response of the target changes over time as $
\bm{h}_{\text{t},k}(t) = \alpha_k\big(\bm{1} - j 2\pi \rho_k \frac{ \bm{f}+ f_{\text{c}}\bm{1}}{c} b(t)\big)$.
Let $\bm{h}_{\text{s},k} = -j2\pi\rho_k\alpha_k/c \cdot \bm{h}_{\text{fb},k}\odot (\bm{f}+f_{\text{c}}\bm{1}) \odot \bm{s}$.
The signal received at antenna $k$ at time $m$ is $\bm{r}_{m,k} = b_{\text{t}}(mT_{\text{rep}})\cdot \bm{h}_{\text{s},k}  + \bm{w}_{m,k} + \text{const}$.
Thus, after removing the constant term we arrive at \eqref{eq:signal_model_kron}.

\begin{figure}
	\centering
	\setlength{\figureheight}{2.5cm}
	\setlength{\figurewidth}{0.7\columnwidth}
	\def\datapath{pgf}
	\input{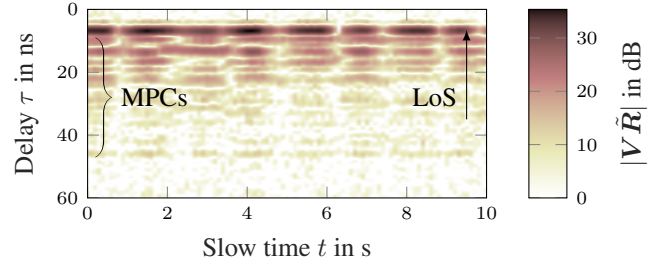}
	\caption{Receive signal $\tilde{\bm{R}}$ of a \gls{uwb} radar measurement of an adult sitting on the back seat of a car.}
	\label{fig:rada_meas}
\end{figure}

Figure \ref{fig:rada_meas} shows a radar measurement from the experiments described in section \ref{sec:performance}. The receive signal is stacked into a matrix $\tilde{\bm{R}} = \big[\tilde{\bm{r}}_0,\,\tilde{\bm{r}}_1,\,\cdots,\,\tilde{\bm{r}}_{M-1}\big]$. To highlight the \glspl{mpc}, the channel is transformed to time domain by the inverse-\gls{fft} matrix $\bm{V}$, where the $[k,n]$-th element of $\bm{V}$ is defined as $\frac{1}{\sqrt{N}} e^{j2\pi nk/N}$.  Each row is approximately a scaled version of the row containing the \gls{los}, as predicted by the derived signal model. The energy in the \glspl{mpc} is used to increase the detection performance.

\vspace{\SecHeadSpacingPre}
\section{Variational Message Passing}
\vspace{\SecHeadSpacingPost}
\label{sec:vmp}

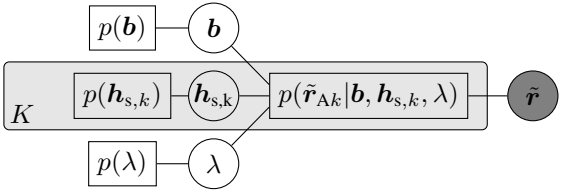
\begin{figure}
	\centering
	\begin{tikzpicture}[every node/.style={draw},
		every circle node/.style={draw,minimum size=.66cm,inner sep=0}]
		\newlength{\nodeheight}
		\setlength{\nodeheight}{0.9cm}
		\newlength{\nodewidth}
		\setlength{\nodewidth}{1.2cm}
		
		\draw[rounded corners=2pt,fill=gray!20!white] (-6.3\nodewidth,-0.5\nodeheight) rectangle (-1\nodewidth,0.5\nodeheight);
		\node[anchor=base,draw=none] at (-6.1\nodewidth,-0.4\nodeheight) {$K$};
		
		\node[style={circle,fill=gray}] at (-0.5\nodewidth,-0\nodeheight) (y) {$\tilde{\bm{r}}$};
		\node at (-2.3\nodewidth,-0\nodeheight) (Py) {$p(\tilde{\bm{r}}_{\text{A}k}|\bm{b},\bm{h}_{\text{s},k},\lambda)$};
		\node[circle] at (-4\nodewidth,-0\nodeheight) (h) {$\bm{h}_\text{s,k}$};
		\node at (-5\nodewidth,-0\nodeheight) (Ph) {$p(\bm{h}_{\text{s},k})$};
		\node[style={circle}] at (-4\nodewidth,-\nodeheight) (lam) {$\lambda$};
		\node at (-5\nodewidth,-\nodeheight) (Plam) {$p(\lambda)$};
		\node[style={circle}] at (-4\nodewidth,\nodeheight) (b) {$\bm{b}$};
		\node at (-5\nodewidth,\nodeheight) (Pb) {$p(\bm{b})$};
		
		\draw (Pb) -- (b) -- ($(Py.west)+(0,1.5mm)$);
		\draw (Ph) -- (h) -- (Py.west);
		\draw (Plam) -- (lam) -- ($(Py.west)-(0,1.5mm)$);
		\draw (y) -- (Py);
		
	\end{tikzpicture}
	\caption{Factor graph of the posterior distribution $p(\bm{b},\bm{h}_{\text{s}},\lambda|\tilde{\bm{r}})$.}
	\label{fig:graph}
\end{figure}

Obtaining the maximum a-posteriori solutions for $\bm{b}$ and $\bm{h}_{\text{s}}$ is computationally infeasible due to the large dimensions of $\bm{b}$ and $\bm{h}_{\text{s}}$. Therefore, we apply a structured mean-field approach to approximate the posterior distribution $p(\bm{b},\bm{h}_{\text{s}},\lambda|\tilde{\bm{r}}) \propto p(\tilde{\bm{r}}|\bm{b},\bm{h}_{\text{s}},\lambda) p(\bm{b}) p(\lambda)\prod_{k=1}^{K}p(\bm{h}_{\text{s},k})$, which is illustrated in Fig. \ref{fig:graph}, with a factorized distribution $q_{1}(\bm{b},\bm{h}_{\text{s}},\lambda)=q_{\text{b}}(\bm{b})q_{\lambda}(\lambda) \prod_{k=1}^{K}q_{\text{h},k}(\bm{h}_{\text{s},k})$ \cite{kirkelund10}. \Gls{vmp} is applied to minimize the Kullbach-Leibler divergence $\mathcal{D}_{\text{KL}}\big(q_{1}\|p(\bm{b},\bm{h}_{\text{s}},\lambda|\tilde{\bm{r}})\big)$ of the true posterior $p(\bm{b},\bm{h}_{\text{s}},\lambda|\tilde{\bm{r}})$ from $q_{1}(\bm{b},\bm{h}_{\text{s}},\lambda)$ by maximizing the \gls{elbo} \cite{winn2005,riegler13,zhang2019,minka2005,blei2017}.
The \gls{elbo} is maximized using coordinate ascent, iteratively maximizing the \gls{elbo} with respect to one distribution $q_{j}\in \mathcal{Q}=\{q_{\text{b}},\,q_{\lambda},\,q_{\text{h},1},\,\cdots,\,q_{\text{h},K}\}$ by
\vspace{-2mm}%
\begin{equation}
	q_j \propto \exp \big\{\big<\ln p(\bm{b},\bm{h}_{\text{s}},\lambda|\tilde{\bm{r}})\big>_{q_{\bar{j}}}\big\}
	\vspace{-2mm}
\end{equation}
while keeping the remaining distributions $q_{\bar{j}} = \prod_{q_k\in \mathcal{Q}\backslash q_j} q_k$ fixed.
Note, that the fixed point can be found analytically, if conjugate priors are used. Therefore, we assume zero-mean Gaussian priors $p(\bm{b})=\normpdf{\bm{b}}{\bm{0}}{\bm{C}_{\text{b}_0}}$ and  $p(\bm{h}_{\text{s},k})=\cnormpdf{\bm{h}_{\text{s},k}}{\bm{0}}{\bm{C}_{\text{h}_0,k}}$ for the respiratory motion and channel vectors, respectively, and Jeffrey's prior $p(\lambda) \propto \lambda^{-1}$ for the noise precision $\lambda$.
The resulting distributions $q_{\text{b}}(\bm{b})=\mathcal{N}(\bm{b}\, |\, \hat{\bm{b}},\,\hat{\bm{C}}_{\text{b}})$, $q_{\text{h},k}(\bm{h}_{\text{s},k})=\mathcal{CN}(\bm{h}_{\text{s},k}\, |\, \hat{\bm{h}}_{\text{s},k},\,\hat{\bm{C}}_{\text{h},k})$, and $q_{\lambda}(\lambda) = \text{Ga}\big(\lambda\, |\, KNM,\, \hat{M}_{\lambda}\big)$ are fully described by the parameters $\hat{\bm{b}}$, $\hat{\bm{C}}_{\text{b}}$, $\hat{\bm{h}}_{\text{s},k}$, $\hat{\bm{C}}_{\text{h},k}$ and $\hat{M}_{\lambda}=KNM/\hat{\lambda}_{1}$.
Let $\hat{E}_{\text{b}}^{[i]}=\trace(\hat{\bm{C}}_{\text{b}}^{[i]}) + \|\hat{\bm{b}}^{[i]}\|^2$ and $\hat{E}_{\text{h}}^{[i]} =\sum_{k=1}^{K}\trace(\hat{\bm{C}}_{\text{h},k}^{[i]}) + \|\hat{\bm{h}}_{\text{s},k}^{[i]}\|^2$, the following messages are computed at iteration $i$:
\begin{align}
\hat{\bm{C}}_{\text{h},k}^{[i]} &= 
\big(\bm{C}_{\text{h}_0,k}^{-1} + \hat{\lambda}_{1}^{[i-1]} \hat{E}_{\text{b}}^{[i-1]} \bm{I}_{N} \big)^{-1}
\label{eq:iter_first} \\
\hat{\bm{h}}_{\text{s},k}^{[i]} &= 
\hat{\lambda}_{1}^{[i-1]} \hat{\bm{C}}_{\text{h},k}^{[i]}\hat{\bm{B}}^{[i-1]\,\hermitian}\tilde{\bm{r}}_{\text{A}k}\\
\hat{\bm{C}}_{\text{b}}^{[i]} &= 
\big(\bm{C}_{\text{b}_0}^{-1} + 2 \hat{\lambda}_{1}^{[i-1]} \hat{E}_{\text{h}}^{[i]} \bm{I}_{L}\big)^{-1} \\
\hat{\bm{b}}^{[i]} &= 
2\hat{\lambda}_{1}^{[i-1]}\hat{\bm{C}}_{\text{b}}^{[i]} \bm{U}^{\text{T}}\real\big\{\hat{\bm{H}}^{[i]\,\hermitian}\tilde{\bm{r}}\big\} \\ \hat{\lambda}_{1}^{[i]} &= \frac{KNM}{\|\tilde{\bm{r}}\|^2 \rmv - \rmv 2\hat{\bm{b}}^{[i]\, \text{T}}\bm{U}^{\text{T}}\real\big\{\hat{\bm{H}}^{[i]\,\text{H}}\tilde{\bm{r}}\big\}+\hat{E}_{\text{b}}^{[i]}\hat{E}_{\text{h}}^{[i]}}
\label{eq:iter_last}
.
\end{align}
After initializing the messages as $\hat{\lambda}_{1}^{[0]}= KNM/\|\tilde{\bm{r}}\|^2$, $\hat{\bm{C}}_{\text{b}}^{[0]}=\bm{C}_{\text{b}_0}$, and $\hat{\bm{b}}^{[0]}$ as a realization drawn from the prior $p(\bm{b})$, equations \eqref{eq:iter_first} trough \eqref{eq:iter_last} are iterated until the messages are converged.
Furthermore, $\hat{\bm{C}}_{\text{b}}^{[i]}$ is calculated by adding a scaled identity matrix to the inverse prior $\bm{C}_{\text{b}_0}^{-1}$. Therefore, the eigenvectors $\bm{U}$ do not change throughout the iterations and can be precomputed based on the chosen prior $\bm{C}_{\text{b}_0}$.

To keep the notation concise, we refrain from explicitly writing iteration indices in the remainder of the paper, referring to the respective values after they are converged.


\vspace{\SecHeadSpacingPre}
\subsection{Detection}

In order to detect the presence of a person, we need to distinguish between two nested models $\mathcal{H}_0$ and $\mathcal{H}_1$, with corresponding likelihoods: $p(\tilde{\bm{r}}|\bm{b}=\bm{0},\bm{h}_{\text{s}}=\bm{0},\lambda,\mathcal{H}_0) = \mathcal{CN}(\tilde{\bm{r}}|\bm{0},\,\lambda^{-1}\bm{I}_{KNM})$ of an empty car, and $p(\tilde{\bm{r}}|\bm{b},\bm{h}_{\text{s}},\lambda,\mathcal{H}_1)=\mathcal{CN}(\tilde{\bm{r}}|\bm{H}\bm{U}\bm{b},\,\lambda^{-1}\bm{I}_{KNM})$ if a person is present. Since the \gls{elbo}, is a lower bound on the logarithmic model evidence $\mathcal{L}(q_j) \leq \ln p(\mathcal{H}_j|\tilde{\bm{r}})$,
we approximate the log odds ratio as $ \ln \frac{p(\mathcal{H}_1|\tilde{\bm{r}})}{p(\mathcal{H}_0|\tilde{\bm{r}})} \approx \mathcal{L}(q_{1}) - \mathcal{L}(q_{0})$, where the \gls{elbo} is given as $\mathcal{L}(q_j) = \big<\ln p(\bm{b},\bm{h}_{\text{s}},\lambda,\mathcal{H}_j|\tilde{\bm{r}})\big>_{q_j(\bm{b},\bm{h}_{\text{s}},\lambda)} + \mathbb{H}(q_j)$ for $j\in\{0,\,1\}$ \cite{winn2005}.
Since $q_{0}(\bm{b},\bm{h}_{\text{s}},\lambda)=q_{0}(\lambda)$ depends only on one parameter, we do not need an iterative update scheme and the \gls{elbo} $\mathcal{L}(q_{0})$ is maximized by $q_0(\lambda)=\text{Ga}\big(\lambda\ |\, KNM,\,\|\tilde{\bm{r}}\|^2\big)$, resulting in $\hat{\lambda}_{0}=KNM/\|\tilde{\bm{r}}\|^2$.
Special considerations must be made regarding the improper prior $p(\lambda)\propto \lambda^{-1}$. Using a proper prior $p(\lambda) = \text{Ga}(\lambda\, |\, d,\,e)$ and taking the limit as $d,e\rightarrow 0$, the test decides for $\mathcal{H}_1$ if
\vspace{-2mm}%
\begin{multline}
	(NM-1)\ln \frac{\hat{\lambda}_{1}}{\hat{\lambda}_{0}} - \sum_{k=1}^{K} \big[\hat{\bm{h}}_{\text{s},k}^\hermitian\bm{C}_{\text{h}_0,k}^{-1}\hat{\bm{h}}_{\text{s},k} + \trace(\bm{C}_{\text{h}_0,k}^{-1}\hat{\bm{C}}_{\text{h},k})\big]  \\
	-\frac{1}{2}\big[\hat{\bm{b}}^\transpose\bm{C}_{\text{b}_0}^{-1}\hat{\bm{b}} + \trace(\bm{C}_{\text{b}_0}^{-1}\hat{\bm{C}}_{\text{b}})\big] + \mathbb{H}(q_{1}) - \mathbb{H}(q_{0}) > \gamma
	\label{eq:detection:decision_final}
	\vspace{-2mm}%
\end{multline}
is larger than the detection threshold $\gamma$.

\vspace{\SecHeadSpacingPre}
\section{Results}
\label{sec:performance}
\vspace{\SecHeadSpacingPost}

To evaluate the performance of the devised algorithm, we consider the case of a single target sitting in a car. A raised-cosine pulse with a bandwidth of $500\,\text{MHz}$ and a roll-off factor of $0.5$ at a centre frequency of $f_{\text{c}}=6.5\,\text{GHz}$ is transmitted every $T_{\text{rep}}=0.1\,\text{s}$ during a measurement duration of $10\,\text{s}$, corresponding to the \gls{uwb} channel 5 in \cite{IEEE:802.15.4}.
The forward and backward channels $\bm{h}_{\text{f}}$ and $\bm{h}_{\text{b}}$ are modeled with an \gls{los} component with power $E_{\text{LoS}}$ at a delay of $\tau_0=1\,\text{m}/c$. The \gls{los} component is followed by a diffuse multipath with exponentially decaying power delay profile with decay constant $\tau_{\text{f}}$. Thus $\bm{h}_{\text{f}}$ and $\bm{h}_{\text{b}}$ are described by the covariance $C_{\text{h}_{\text{f}}}[n,n^\prime]= C_{\text{h}_{\text{b}}}[n,n^\prime] = \big[E_{\text{LoS}} + E_{\text{DM}} \left.\big(1+j2\pi\tau_{\text{f}}\Delta f (n-n^\prime)\big)^{-1}\big]\right. e^{-j2\pi\tau_0\Delta f (n-n^\prime)}$.
We choose $\tau_{\text{f}}=20\,\text{ns}$ and $K_{\text{LoS}}=\frac{E_{\text{LoS}}}{E_{\text{DM}}}=0.75$, since these values were observed by test measurements.
If the delay of the \gls{los} component is known, the prior covariance $\bm{C}_{\text{h}_0}$ can be calculated from $C_{\text{h}_{\text{f}}}[n,n^\prime]$. However, we do not assume the distance to the target to be known a-priory.
Therefore, we choose the covariance for the forward-backward channel to follow the shape of a gamma distribution $C_{\text{fb}}[n,n^\prime] \propto \big(1+j2\pi \tau_{\text{f}} \Delta f (n-n^\prime)\big)^{-a}$. We select $a=1+\frac{2\tau_0}{\tau_{\text{f}}}$ such that the peak (in time domain) corresponds to the expected target distance.
This choice of prior enhances the detection performance compared to a flat prior by incorporating the decay constant of the channel without requiring the exact target distance to be known.
The typical respiratory frequency for an adult is between 9 and 21 breaths per minute \cite{fairclough2012}, where as the respiratory frequency of babies can go as high as 60 breaths per minute\cite{frey98}. Therefore,
we assume a prior covariance $\bm{C}_{\text{b}_{\text{t}}}$ with a rectangular double sided power spectral density $S_{\text{b}}(f) = \frac{1}{2(f_{\text{b,max}}-f_{\text{b,min}})}$ for $f_{\text{b,min}} \leq |f| \leq f_{\text{b,max}}$ and $0$ elsewhere, and select $f_{\text{b,min}} = 9/60 \,\text{Hz}$ and $f_{\text{b,max}}=1\,\text{Hz}$.

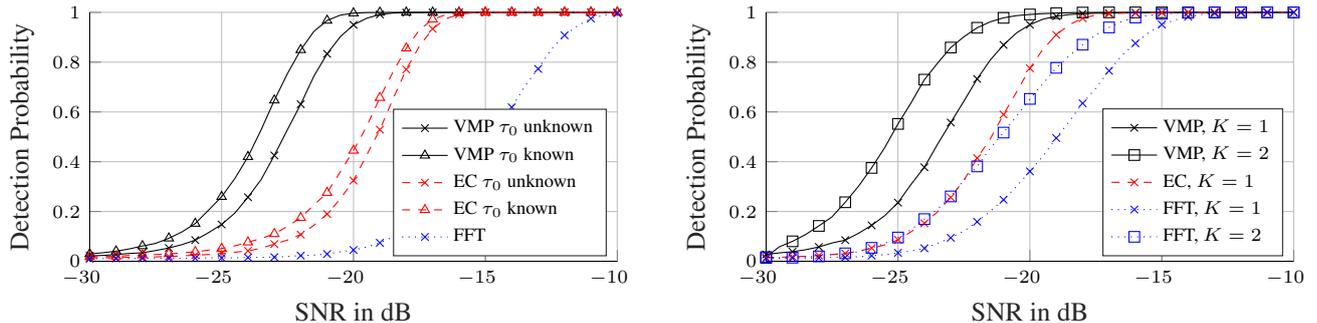
\begin{figure*}
	\centering
	\setlength{\figureheight}{3.3cm}
	\setlength{\figurewidth}{0.39\textwidth}
	\def\datapath{pgf}
%
%
 
 
\pgfplotsset{every axis/.append style={
  label style={font=\scriptsize},
  legend style={font=\scriptsize},
  tick label style={font=\scriptsize},
  xticklabel={
    \ifdim \tick pt < 0pt
      \pgfmathparse{abs(\tick)}%
      \llap{$-{}$}\pgfmathprintnumber{\pgfmathresult}
   \else
      \pgfmathprintnumber{\tick}
   \fi
}}}
 
\begin{tikzpicture}

\begin{axis}[%
	width=\figurewidth,
	height=\figureheight,
	at={(0\figurewidth,0\figureheight)},
	scale only axis,
	xmin=-30,
	xmax=-10,
	xlabel style={font=\color{white!15!black}},
	xlabel={SNR in dB},
	ymin=0,
	ymax=1,
	ylabel style={font=\color{white!15!black}},
	ylabel={Detection Probability},
	axis background/.style={fill=white},
	axis x line*=bottom,
	axis y line*=left,
	xmajorgrids,
	ymajorgrids,
	legend style={at={(0.98,0.02)}, anchor=south east, legend cell align=left, align=left, draw=white!15!black},
	]

	\addplot [color=black, solid, mark=x, mark options={solid, black}, mark repeat=2]
	table[]{\datapath/detectionResults-elbo-exp.tsv};
	\addlegendentry{VMP $\tau_0$ unknown}
	
	\addplot [color=black, solid, mark=triangle, mark options={solid, black}, mark repeat=2]
	table[]{\datapath/detectionResults-elbo-ideal.tsv};
	\addlegendentry{VMP $\tau_0$ known}
	
	\addplot [color=red, dashed, mark=x, mark options={solid, red}, mark repeat=2]
	table[]{\datapath/detectionResults-ec-exp.tsv};
	\addlegendentry{EC $\tau_0$ unknown}
	
	\addplot [color=red, dashed, mark=triangle, mark options={solid, red}, mark repeat=2]
	table[]{\datapath/detectionResults-ec-ideal.tsv};
	\addlegendentry{EC $\tau_0$ known}
	
	\addplot [color=blue, dotted, mark=x, mark options={solid, blue}, mark repeat=2] 
	table[]{\datapath/detectionResults-fft.tsv};
	\addlegendentry{FFT}
	
\end{axis}

\begin{axis}[%
	width=\figurewidth,
	height=\figureheight,
	at={(0.5\textwidth,0\figureheight)},
	scale only axis,
	xmin=-30,
	xmax=-10,
	xlabel style={font=\color{white!15!black}},
	xlabel={SNR in dB},
	ymin=0,
	ymax=1,
	ylabel style={font=\color{white!15!black}},
	ylabel={Detection Probability},
	axis background/.style={fill=white},
	axis x line*=bottom,
	axis y line*=left,
	xmajorgrids,
	ymajorgrids,
	legend style={at={(0.98,0.02)}, anchor=south east,legend cell align=left, align=left, draw=white!15!black},
	]
	\addplot [color=black,mark=x, mark options={solid, black}, mark repeat=2]
	table[]{\datapath/detectionResultsMeas-elbo.tsv};
	\addlegendentry{VMP, $K=1$}
	
	\addplot [color=black,mark=square, mark options={solid, black}, mark repeat=2]
	table[]{\datapath/detectionResultsMeasMultistatic-elbo.tsv};
	\addlegendentry{VMP, $K=2$}
	
	\addplot [color=red, dashed, mark=x, mark options={solid, red}, mark repeat=2]
	table[]{\datapath/detectionResultsMeas-ec.tsv};
	\addlegendentry{EC, $K=1$}
	
	\addplot [color=blue, dotted, mark=x, mark options={solid, blue}, mark repeat=2]
	table[]{\datapath/detectionResultsMeas-fft.tsv};
	\addlegendentry{FFT, $K=1$}
	
	\addplot [color=blue, dotted, mark=square, mark options={solid, blue}, mark repeat=2]
	table[]{\datapath/detectionResultsMeasMultistatic-fft.tsv};
	\addlegendentry{FFT, $K=2$}
	
\end{axis}

\end{tikzpicture}%
	\caption{Detection performance of the \acrshort{vmp}-based detector compared to an estimator-correlator (EC) and \acrshort{fft} detector on simulated data (left) and measurements (right).}
	\label{fig:results}
\end{figure*}

To evaluate the performance, a Monte-Carlo simulation with $10^5$ runs was performed at each $\text{SNR}=\frac{\lambda\|\bm{b}\|^2 \|\bm{h}_\text{s}\|^2}{NM}$ from $-30$\,dB to $10$\,dB in $0.5$\,dB steps considering a monostatic setup. The threshold for all detectors was set such that a constant false alarm rate of $p_\text{FA}=0.01$ was achieved at each \gls{snr} value.
As comparison, we evaluated the detection performance of an estimator-correlator, which models $\tilde{\bm{r}}$ as a Gaussian process to incorporate \glspl{mpc} into the detection \cite{moederl22}, and an \gls{fft}-based detector which computes the \gls{fft} over the rows of the matrix $\bm{V}\tilde{\bm{R}}$ and compares the peak against a threshold. For the \gls{fft}-based approach the channel is transformed to the time domain to concentrate the signal energy in the delay bin corresponding to the \gls{los} component for easier detection. However, no \glspl{mpc} are incorporated in the detection. 
The results are depicted in Fig. \ref{fig:results}. The proposed \gls{vmp}-based algorithm has a better detection rate compared to the two other methods, even when using the modified prior which does not require a-priory knowledge of the distance to the target.
Specifically, at an \gls{snr} of $-20$\,dB, the \gls{vmp}-based detection achieves a detection rate of approximately $0.95$ while the estimator-correlator achieves $0.32$ and the \gls{fft} detector achieves $0.05$ in case the target distance is not known.
This difference in performance can be explained by the different level of incorporation of the signal model into the detection:
The \gls{vmp}-based detector incorporates the signal model more rigidly in the detection compared to the estimator-correlator which only accounts for the correlation between different columns of the matrix $\tilde{\bm{R}}$, or the \gls{fft}-based detector which does not account for \glspl{mpc} at all.

A measurement campaign including 34 participants (6 female and 28 male) was performed during which 177 minutes of data with a sample rate of $T_{\text{rep}}=0.1\,\text{s}$ have been collected using a multistatic setup with one transmit and two receive antennas. The participants were instructed to sit motionless in the car while breathing normally. The measurements were performed using an M-sequence channel sounder in a Seat Leon and a Citroen Picasso. The measurement equipment including channel sounder, cables and connectors has been calibrated before the measurement and the same transmit pulse was used as in the simulations.
The data was split in to non-overlapping chunks of $10\,\text{s}$ length, for a total of 481 measurements.
For each \gls{snr} value, $100$ independent noise realizations were added. The resulting detection performance is shown in Fig. \ref{fig:results} for either both receive antennas ($K=2$) or considering only one receive antenna ($K=1$).
The \gls{fft}-based approach performs significantly better on the measured data compared to the simulated data, since the respiratory motion of adults is closer to a periodic signal and has less randomness than the Gaussian process used in the simulations.
Although the performance difference between the \gls{vmp}-based approach and the other methods is less than in the simulations, the best detection probability is still achieved by the \gls{vmp}-based detector.


\vspace{\SecHeadSpacingPre}
\section{Conclusion}
\label{sec:conclusion}
\vspace{\SecHeadSpacingPost}

We present a novel \gls{vmp}-based approach to detect the presence of a person by their respiratory chest motion using \gls{uwb} radar signals. The devised algorithm significantly outperforms the comparison methods on simulated as well as measured data. The superior performance of the presented detection algorithm is achieved trough the use of \glspl{mpc}, which are shown to carry a significant amount of signal energy in small spaces, such as the interior of a car.
Furthermore, deviations from a strictly periodic respiration pattern, such as pauses with different length in between breaths do not impact the performance of the devised algorithm, since the respiratory motion $b_{\text{t}}(t)$ is modeled as a random process and not as a periodic function as e.g. in the FFT-based detection approach.
Additionally, the devised algorithm can be used for contact-free vital sign estimation since the respiratory chest motion is estimated as part of the algorithm. However, the received signal $\tilde{\bm{r}}$ depends on the product of $\bm{b}$ and $\bm{h}_{\text{s}}$ which results in an ambiguity in the sign of $\hat{\bm{b}}$ and $\hat{\bm{h}}_{\text{s}}$.

Since many future cars will be equipped with \gls{uwb} nodes, e.g. as part of the keyless entry system, the developed algorithm provides occupancy sensing capabilities to these cars without the need and increased manufacturing costs of dedicated sensors.


\bibliographystyle{IEEEtran}
\bibliography{IEEEabrv,references_moederlICASSP2023}


%
\end{document}